\documentclass[preprint,12pt]{elsarticle}

\usepackage{amssymb, amsmath, gensymb, caption, subcaption, float}

\journal{Proceedings of the Royal Society A}

\begin{document}

\begin{frontmatter}

\title{Determinants of successful mitigation in coupled social-climate dynamics}

\author[inst1]{Longmei Shu}

\affiliation[inst1]{organization={Department of Mathematics},
            addressline={27 N. Main Street}, 
            city={Hanover},
            postcode={03755}, 
            state={NH},
            country={USA}}

\author[inst1,inst2]{Feng Fu}

\affiliation[inst2]{organization={Department of Biomedical Data Science},
            addressline={Geisel School of Medicine at Dartmouth}, 
            city={Lebanon},
            postcode={03756}, 
            state={NH},
            country={USA}}

\begin{abstract}
Understanding the impact of human behavior is crucial for successful mitigation of climate change across the globe. To shed light onto this issue, here we couple the forest dieback model with human behaviors. Using evolutionary game theory, we build a time-delay system where forest growth is impacted by both temperature and human mitigation choices, the latter being informed by temperature forecasts. Simulations of the coupled system over 200 years show us the varying outcomes: forest dies out and no one is a mitigator, forest dies out and everyone is a mitigator,  or the forest survives and everyone is a mitigator. There exist rare cases where no one is a mitigator and yet the forest survives, but with a low coverage. We also find occasional oscillations where the proportion of mitigators vary between 0 and 1. Our results are based on simple models but have profound insights into determinants of behavior changes desired in social-climate dynamics.
\end{abstract}


\end{frontmatter}

\section{Introduction}

Climate change, or global warming is one of many public goods games where social cooperation can play a very important role in averting potential catastrophic scenarios~\cite{kemp2022climate, santos2021dynamics, levin2021governance, kruczkiewicz2021compound}. Typically such collective-risk social dilemmas have complicated dynamics \cite{wang2009emergence, zhang2013tale, wu2013increased,tilman2021evolution}. Experiments with real people and money reward were also carried out to see how people would act in threshold climate games \cite{milinski2008collective,jacquet2013intra}. Field experiments in multiple countries around the world \cite{cardenas2017fragility} give us some insights on the climate problem as well. 

Models increasingly help us understand the interactions between the carbon cycle, the climate system, human processes, and the impact of policies \cite{moore22}. To ensure those policy decisions are robust to uncertainties, multiple scenarios are often laid out, ranging from carbon emission trajectories to socio-economic systems pathways. Influence in these models usually flows in one-direction, from socio-economic systems to the Earth system. Yet, two-way feedback mechanisms link climate and social processes: human behavior changes the climate, and the climate changes human opinions and consequently human behavior. Coupled human-environment models are already widely applied to study other systems such as fisheries and forests and the need for coupled social-climate models has been noted \cite{laborde16,beckage20,beckage2022incorporating}. Such eco-evolutionary models where the environment changes depending on the actions of the players again adds to the rich dynamics of the system \cite{wang2020eco, shu2022eco}.

Many social-climate models are complicated and we don't have a good intuition of the dynamics over time. Here we focus on a simple climate model, the forest dieback model, and couple it with a replicator equation that determines the percentage of mitigators in the population based on the warming over the past 10 years. So the social part of our model has a time-delay effect, and the growth rate of the forest depends on the percentage of mitigators in the population. We study the behavior of this coupled social-climate model under changing parameter values, the ambient temperature (or background temperature of the environment), and a sensitivity to warming. The sensitivity to warming decides how likely people are going to adopt mitigative behavior based on current warming. We solve our delay system over 200 years and see if the forest dies out or survives, and how much of the population mitigates climate change.

Since our model is relatively simple, we are able to analyze the equilibrium points well and see how they change when the parameters change. We run simulations over a wide range of parameters to see all possible rich dynamics of our model. Our findings, though derived from simple models, offer deep understanding into the factors influencing mitigation choices in the context of social-climate interactions.

\section{Methods and Model}
\subsection{Forest dieback model}

We consider the climate and temperature for Amazonian forests. The vegetation coverage $v$ is between 0 and 1, with 0 for bare soil and 1 for full forest coverage. Then $v$ satisfies the following equation
\begin{align}
    \frac{dv}{dt}=gv(1-v)-\gamma v.
    \label{forest}
\end{align}
Here $\gamma=0.2$ is the disturbance rate \cite{ritchie21} and $g$ is the growth rate. The equilibrium points of the system have to satisfy
\begin{align*}
    gv(1-v)=\gamma v,\quad v=0\text{ or }v=1-\frac{\gamma}{g}.
\end{align*}

\begin{figure}[h]
    \centering
     \begin{subfigure}[b]{0.45\textwidth}
         \includegraphics[scale=0.5]{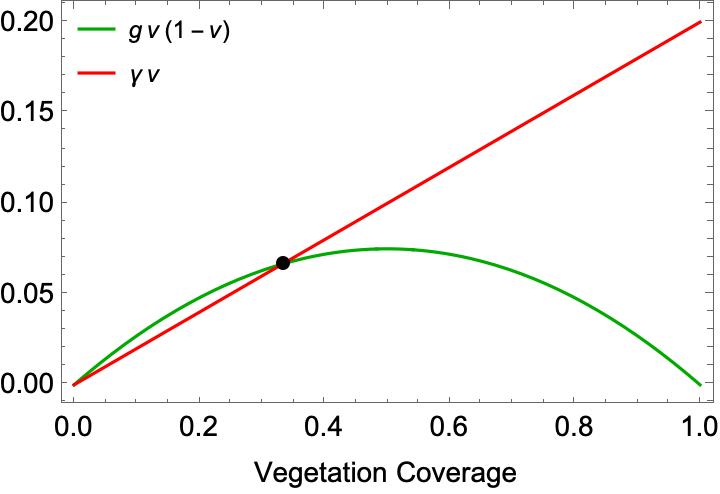}
         \caption{$g=0.3$}
     \end{subfigure}
     \qquad
     \begin{subfigure}[b]{0.45\textwidth}
         \includegraphics[scale=0.5]{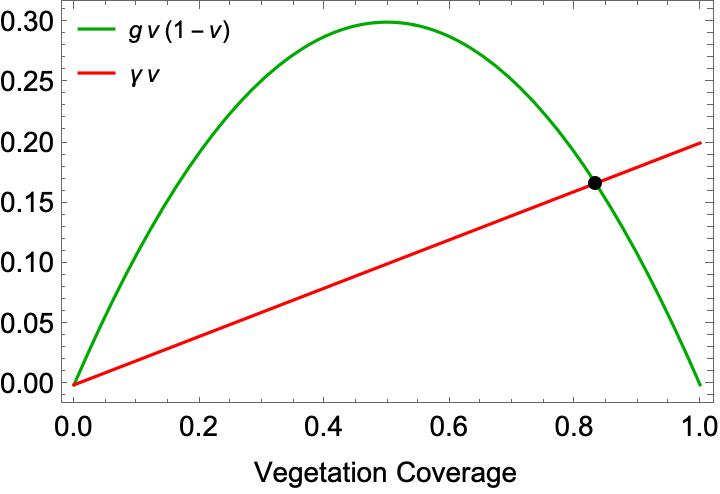}
         \caption{$g=1.2$}
     \end{subfigure}\\
     \begin{subfigure}[b]{0.45\textwidth}
        \includegraphics[scale=0.5]{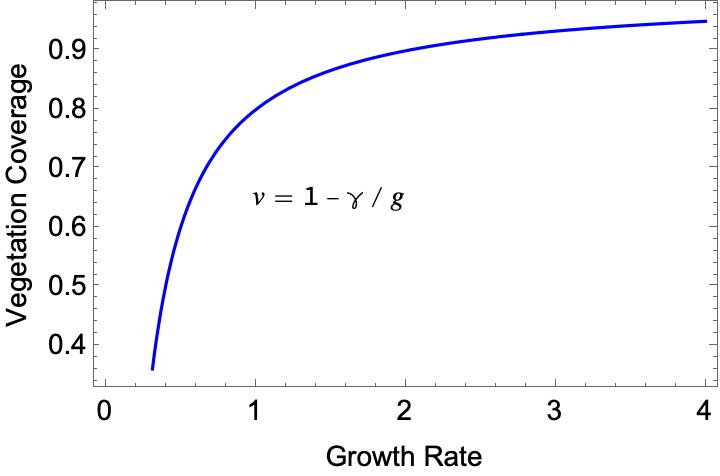}
        \caption{$v=1-\gamma/g$}
     \end{subfigure}
     \caption{Equilibrium under varying growth rate $g$}
     \label{g}
\end{figure}

The fixed point under varying values of $g$ are plotted in Figure \ref{g}. We can see the vegetation coverage has a positive robust stable fixed point when $g>2$.

In the forest dieback model $g$ is given by
\begin{align}
g=g_0\left[1-\left(\frac{T-T_{opt}}{\beta}\right)^2\right].
\label{growth}
\end{align}
Here $g_0=2$ is the maximum growth rate, and $\beta=10$ is the half-width of the growth versus temperature curve. $T_{opt}=28^{\circ}$C represents the optimal temperature for plant growth.

$T$ is the actual temperature and is given by
\begin{align}
T=T_v+(1-v)a.
\label{ambient}
\end{align}
Here $a=5$ is the difference between surface temperature of bare soil and forest and $T_v$ is the temperature with full forest coverage. For convenience, we will call it the ambient temperature from here on. 

Plug \eqref{growth} and \eqref{ambient} into \eqref{forest}, we get fixed points of the forest dieback model satisfies
\begin{align}
v\left\{2(1-v)\left[1-\left(\frac{T_v+a(1-v)-T_{opt}}{\beta}\right)^2\right]-\gamma\right\}=0.
\label{fixed-points}
\end{align}

Plug in all the parameter values, we get 
\begin{align*}
    v=0\text{ or }(1-v)[1-0.01(T_v-23-5v)^2]=0.1.
\end{align*}

\begin{figure}[h]
\centering
\includegraphics[scale=0.6]{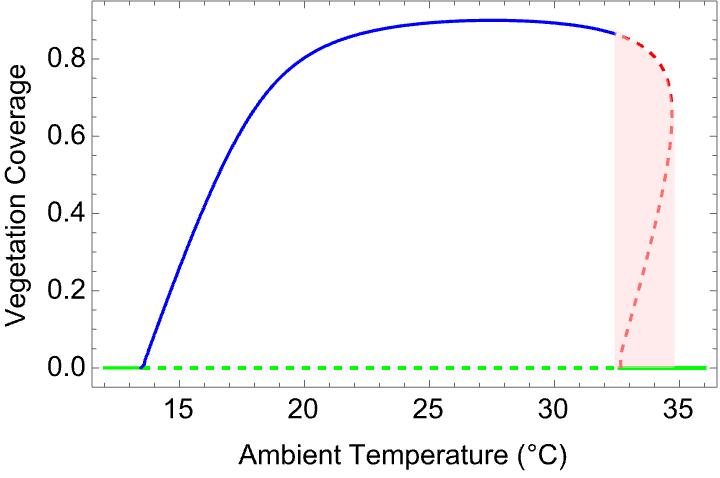}
\caption{Equilibrium under varying ambient temperature}
\label{Tv}
\end{figure}

Equilibrium points for varying values of $T_v$ are shown in Figure \ref{Tv}. We can see bifurcation \cite{may1976simple, wang2009emergence, liu2023coevolutionary} happens and we have two different fixed points when the ambient temperature $T_v$ is between 32.5$^\circ$C and 34.7$^\circ$C. The green dashed line represents the unstable fixed points $v=0$, the blue curve represents the stable and robust fixed points, and the red dashed curve represents fixed points in the shaded bifurcation region.

\subsection{Human actions}
We divide the population into two groups, mitigators (M) and nonmitigators (N). And the percentage of mitigators is $x$, the percentage of nonmitigators is $y$. So $0<x,y<1, x+y=1$. The payoffs of mitigators and nonmitigators are
\begin{align*}
E_M &=-\alpha+\frac{1}{2}f(T_f)+\delta x, \\
E_N &=-\frac{1}{2}f(T_f)+\delta y.
\end{align*}
Here $\alpha=1$ is the cost of mitigation,  and $\delta=1$ is the strength of social norms. \cite{menard21}

Let $\bar E$ be the average payoff in the population, then the proportion of mitigators satisfies the following replicator equation.
\begin{align*}
\frac{dx}{dt}&=x(E_M-\bar E)\\
&=x(E_M-xE_m-yE_N)\\
&=x(1-x)(E_M-E_N)\\
&=x(1-x)[-\alpha+f(T_f)+\delta(2x-1)]
\end{align*}
So far we haven't explained what $f(T_f)$ is yet. This term corresponds to the cost of global warming and satisfies the following equation.
\begin{align}
f(T_f)=\frac{f_{max}}{1+e^{-w(T_f-T_c)}}
\label{warming}
\end{align}
Here $f_{max}=5$ is the maximum warming cost, $w=3$ is the non-linearity of warming cost, and $T_c$ is the critical temperature \cite{menard21}. $T_f$ is the perceived temperature rise given by
\begin{align}
    T_f(t)=\frac{t_f}{t_p}[T(t)-T(t-t_p)]
    \label{perception}
\end{align}
Here $t_p=10$ is the number of previous years used for temperature projection, $t_f=15$ is the number of years ahead for temperature projection.

Plug \eqref{warming},\eqref{perception} in the replicator equation, we get
\begin{align}
    \frac{dx}{dt}&=x(1-x)\left[\delta(2x-1)-\alpha+\frac{f_{max}}{1+e^{-w[t_f/t_pT(t)-t_f/t_pT(t-t_p)-T_c]}}\right].
    \label{mitigator}
\end{align}

\subsection{Coupled social-climate model}

In the previous section, we already considered how the temperature affects human actions. Now we will try to consider the effects of human actions on the climate as well. We will modify the growth rate given by \eqref{growth} by adding a term that depends on $x$, the proportion of mitigators,

\begin{align*}
g=g_0\left[1-\left(\frac{T-T_{opt}}{\beta}\right)^2\right]\eta(x).
\end{align*}

Obviously we want $\eta(x)$ to increase when $x$ increases, since the more mitigators there are, the better the forest grows. To make sure that the change in the proportion of mitigators has an observable effect in the forest coverage, we set $\eta(x)=0.2+0.4x$ so that when $x$ changes from 0 to 1, roughly $g$ changes between 0.4 and 1.2, where the equilibrium forest coverage is sensitive to the growth rate, see Figure \ref{g}.

Plug in the values for all parameters other than $T_v$ and $T_c$, the two equations we are coupling are

\begin{equation}
\begin{cases}
\displaystyle
\frac{dv}{dt}&=2[1-0.01(T_v-23-5v)^2](0.2+0.4x)v(1-v)-0.2v,\\
\vspace{5pt}\\
\displaystyle
\frac{dx}{dt}&=x(1-x)\left[2x-2+\dfrac{5}{1+e^{-3[-7.5v(t)+7.5v(t-10)-T_c]}}\right].
\label{social-couple} 
\end{cases}
\end{equation}

This is a system of autonomous nonlinear delay differential equations.

\section{Theoretical Analysis}
We now turn to analyze the coupled social-climate model in Eq.~\ref{social-couple}. Let's look at the fixed points of this system. For $dv/dt=0$ we get either $v=0$ or 
\begin{align}
    h(v,x,T_v)=2(1-v)(0.2+0.4x)[1-0.01(T_v-23-5v)^2]-0.2=0.
    \label{root}
\end{align}

\begin{figure}[h]
    \centering
     \begin{subfigure}[b]{0.45\textwidth}
         \includegraphics[scale=0.5]{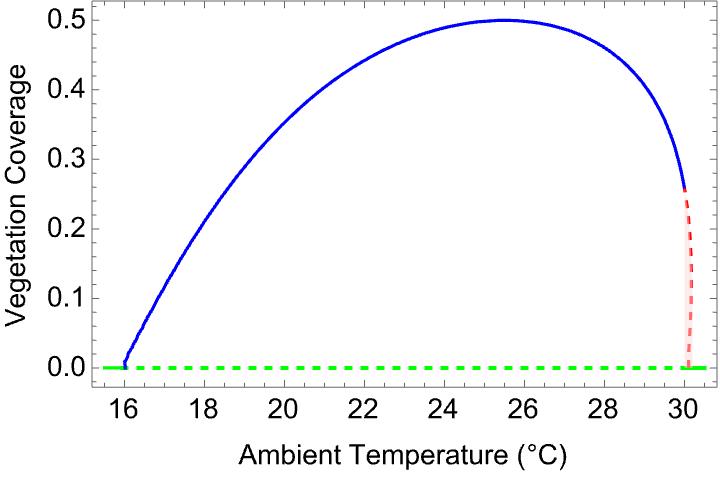}
         \caption{$x=0$}
     \end{subfigure}
     \qquad
     \begin{subfigure}[b]{0.45\textwidth}
         \includegraphics[scale=0.5]{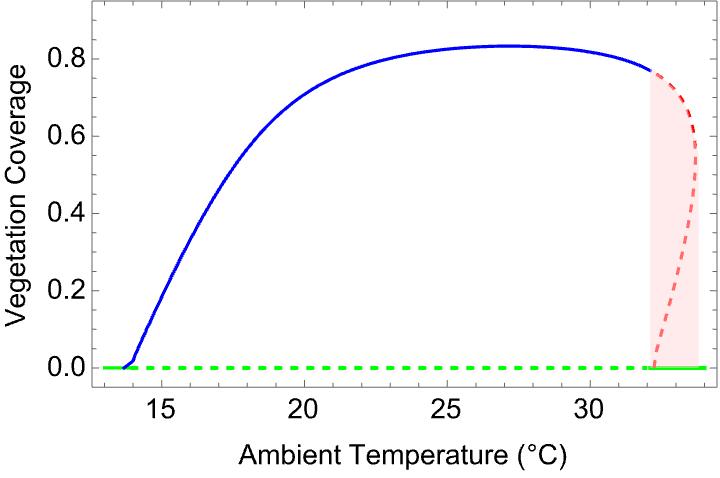}
         \caption{$x=1$}
     \end{subfigure}
     \caption{Equilibrium under varying ambient temperature and mitigation}
     \label{Tv-x}
\end{figure}
\vspace{-5pt}
$h(v,x,T_v)$ depends on the value of $x$ smoothly. As we vary the value of $x$ through $[0,1]$, we expect the curve defined by \eqref{root} for $T_v$ and $v$ to change smoothly. We show the curves for $x=0$ and $x=1$ in Figure \ref{Tv-x}. As one can see, the solution curve is well defined for $v$ in $(0,1)$ for a given value of $T_v$ up to a certain point, beyond which bifurcation happens where one ambient temperature $T_v$ corresponds to two different values for $v$ and a small change in $T_v$ can result in a dramatic change in $v$ value. For $x=0$, bifurcation happens when $T_v$ is between 30$^\circ$C and 30.2$^\circ$C. For $x=1$, bifurcation happens when $T_v$ is between 32.1$^\circ$C and 33.7$^\circ$C.

One can check that above the blue curve $h(v,x,T_v)<0$ and below the blue curve $h(v,x,T_v)>0$, so $v=0$ is unstable and the fixed point determined by the blue curve is stable and robust.

\begin{figure}[H]
    \centering
     \begin{subfigure}[b]{0.45\textwidth}
         \includegraphics[scale=0.5]{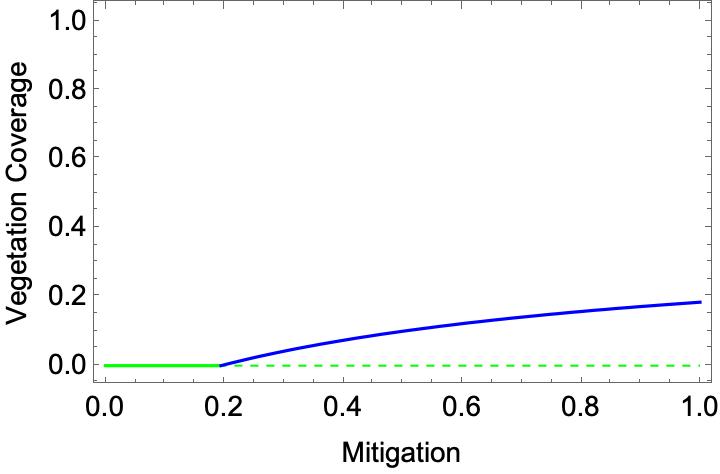}
         \caption{$T_v$=15$^\circ$C}
     \end{subfigure}
     \qquad
     \begin{subfigure}[b]{0.45\textwidth}
         \includegraphics[scale=0.5]{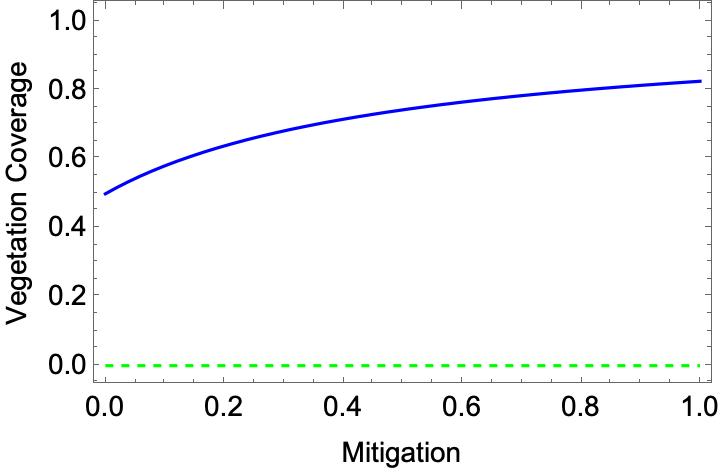}
         \caption{$T_v$=25$^\circ$C}
     \end{subfigure}\\
     \begin{subfigure}[b]{0.45\textwidth}
         \includegraphics[scale=0.5]{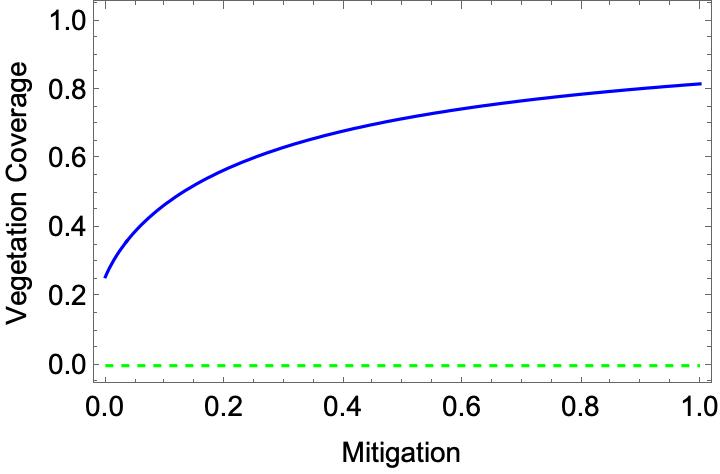}
         \caption{$T_v$=30$^\circ$C}
     \end{subfigure}
     \qquad
     \begin{subfigure}[b]{0.45\textwidth}
         \includegraphics[scale=0.5]{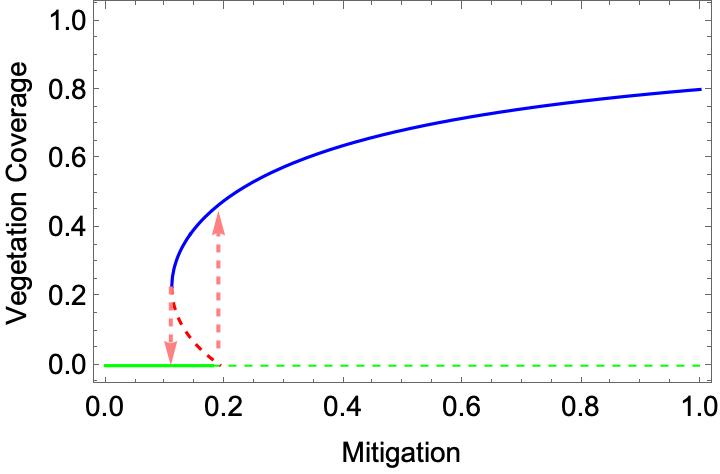}
         \caption{$T_v$=31$^\circ$C}
    \end{subfigure}\\
    \begin{subfigure}[b]{0.45\textwidth}
         \includegraphics[scale=0.5]{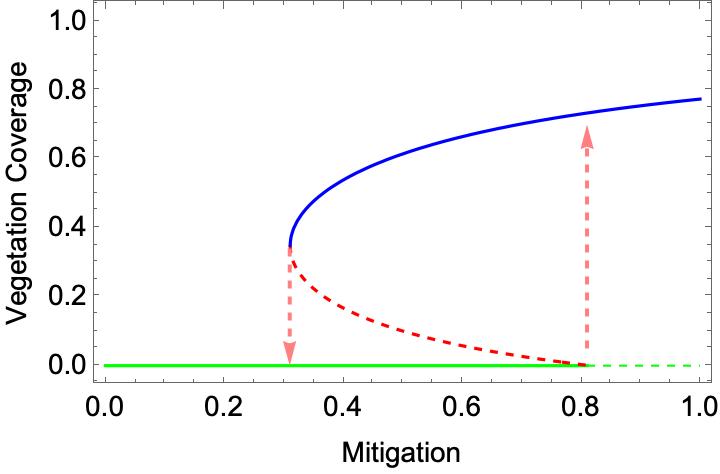}
         \caption{$T_v$=32$^\circ$C}
     \end{subfigure}
     \qquad
     \begin{subfigure}[b]{0.45\textwidth}
         \includegraphics[scale=0.5]{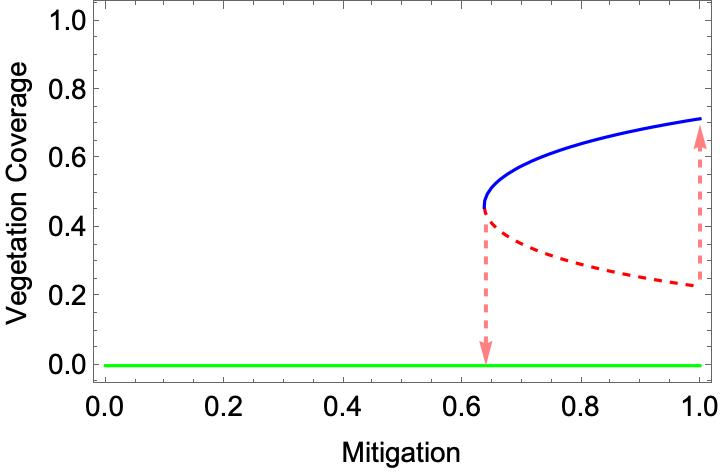}
         \caption{$T_v$=33$^\circ$C}
     \end{subfigure}
     \caption{Equilibrium under varying mitigation and ambient temperature}
     \label{v-x-Tv}
\end{figure}

To have a better understanding of our equation, we also show plots for the equilibrium points under different fixed ambient temperatures. We still expect the relationship curve to change smoothly as the ambient temperature changes.  For ambient temperatures between 18$^\circ$C and 30$^\circ$C, the curve looks similar to the case where $T_v=25^\circ$C in figure \ref{v-x-Tv}. Below this curve, $h(v,x,T_v)>0$, and above the curve $h(v,x,T_v)<0$. As the ambient temperature lowers, this curve shifts down until it disappears from our phase space around 14$^\circ$C. However, as the ambient temperature rises, the curve doesn't rise up and disappear. Instead we see a horizontal parabola-like curve that moves up and right until it disappears from the right side of our phase space around 34$^\circ$C. To show this process, we plot the curve under ambient temperatures 30$^\circ$C, 31$^\circ$C, 32$^\circ$C, and 33$^\circ$C in Figure \ref{v-x-Tv}.

As one can see, when the temperature is higher than 30$^\circ$C, inside of the parabola-like curve, $dv/dt>0$, and outside of the parabola-like curve $dv/dt<0$. This gives us a hysteresis loop and backward bifurcation. The blue curve represents stable fixed points and the red dashed curve represents unstable fixed points even though they are technically the same curve, i.e. they are determined by the same equation. Below a certain vegetation coverage, the fixed point is unstable and above a certain vegetation coverage, the fixed point is stable.

Notice that the green straight line $v=0$ sometimes is dashed and sometimes is solid, the solid parts correspond to stable fixed points and dashed parts correspond to unstable fixed points.

For $dx/dt=0$ we get $x=0$, $x=1$, or

\begin{align}
    x^*(T_c)=1-\frac{2.5}{1+e^{3T_c}}.
    \label{x}
\end{align}

\begin{figure}[h]
\centering
\includegraphics[scale=0.6]{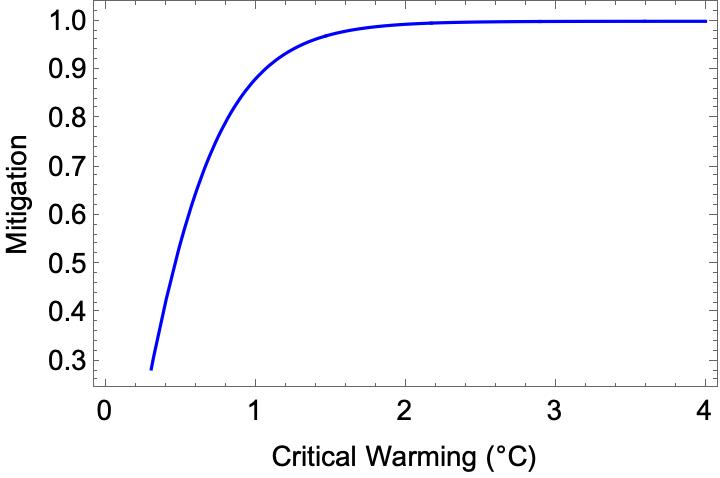}
\caption{Equilibrium under varying critical warming}
\label{Tc-x}
\end{figure}

For $x^*$ to be between 0 and 1, we need $T_c>1/3\ln(1.5)\approx0.135^\circ C$. And for the default value listed in \cite{ritchie21}, where $T_c=2.5^\circ C$, we have $x^*(2.5)\approx0.9986$, very close to $1$. Among the three fixed point values, $x=0$ and $x=1$ are stable and $x^*(T_c)$ is unstable. We show a graph of \eqref{x} in Figure \ref{Tc-x}. As one can see $x^*$ approaches 1 as soon as the critical warming reaches around 2$^\circ$C.

\begin{figure}[H]
    \centering
     \begin{subfigure}[b]{0.45\textwidth}
         \includegraphics[scale=0.45]{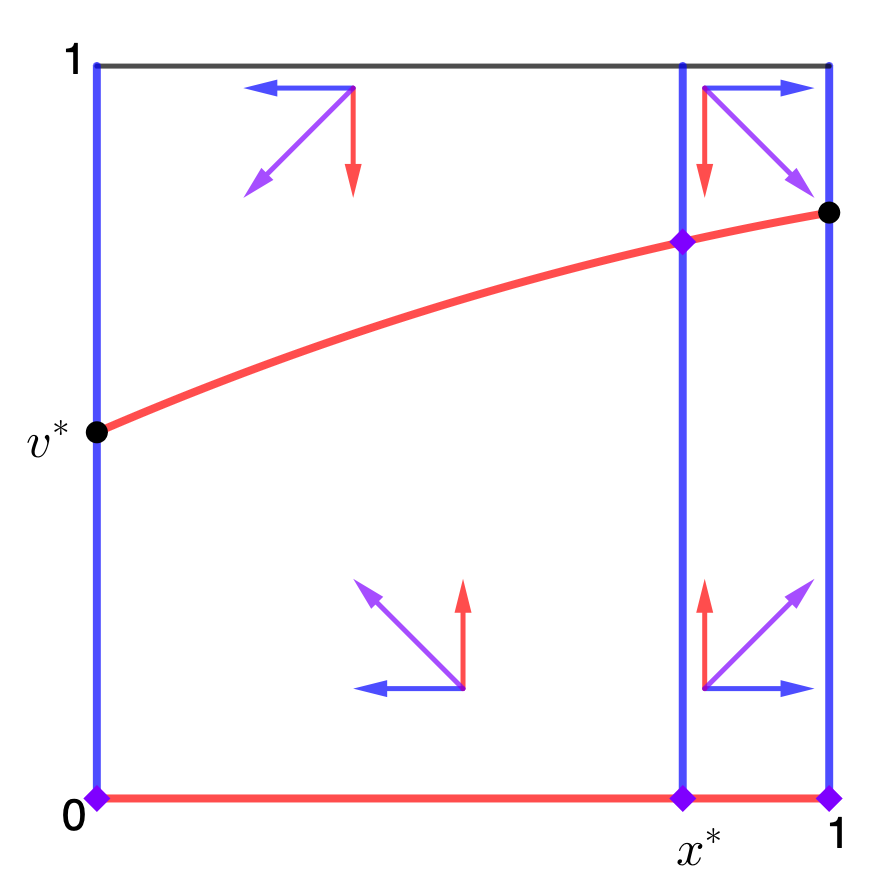}
         \caption{6 fixed points}
     \end{subfigure}
     \qquad
     \begin{subfigure}[b]{0.45\textwidth}
         \includegraphics[scale=0.4]{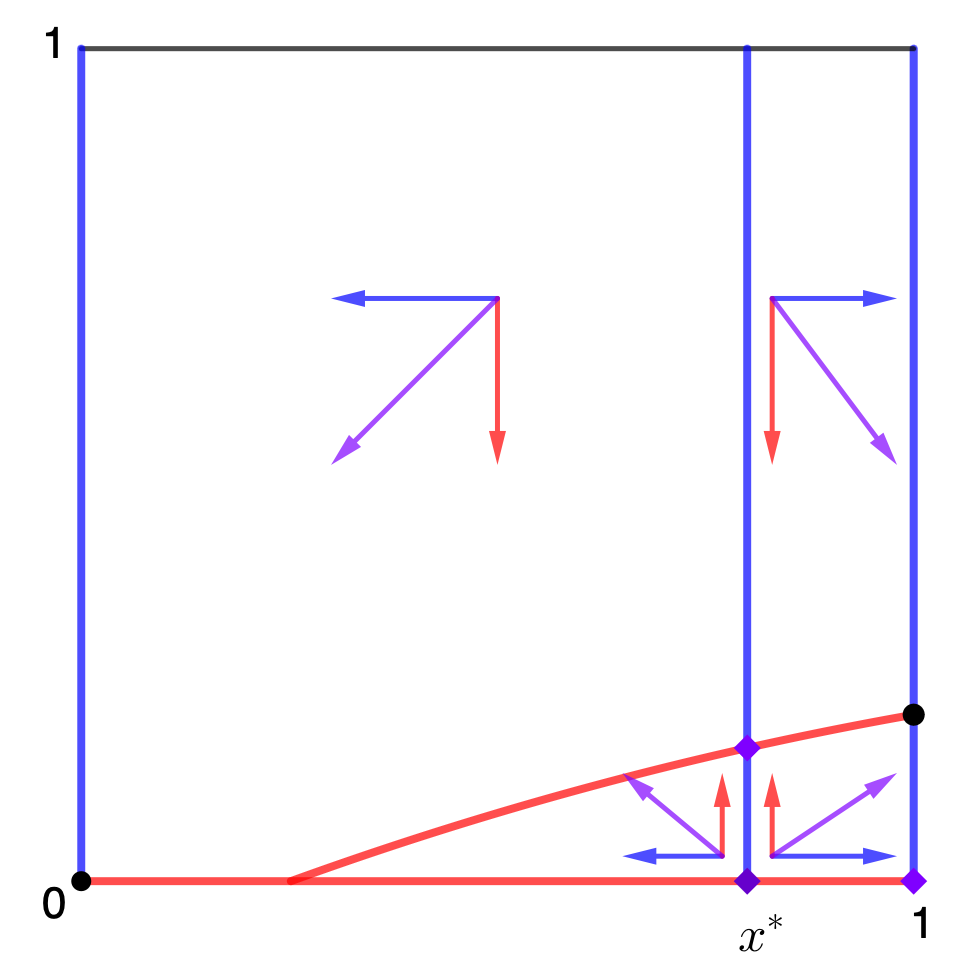}
         \caption{5 fixed points}
     \end{subfigure}\\
     \begin{subfigure}[b]{0.45\textwidth}
         \includegraphics[scale=0.4]{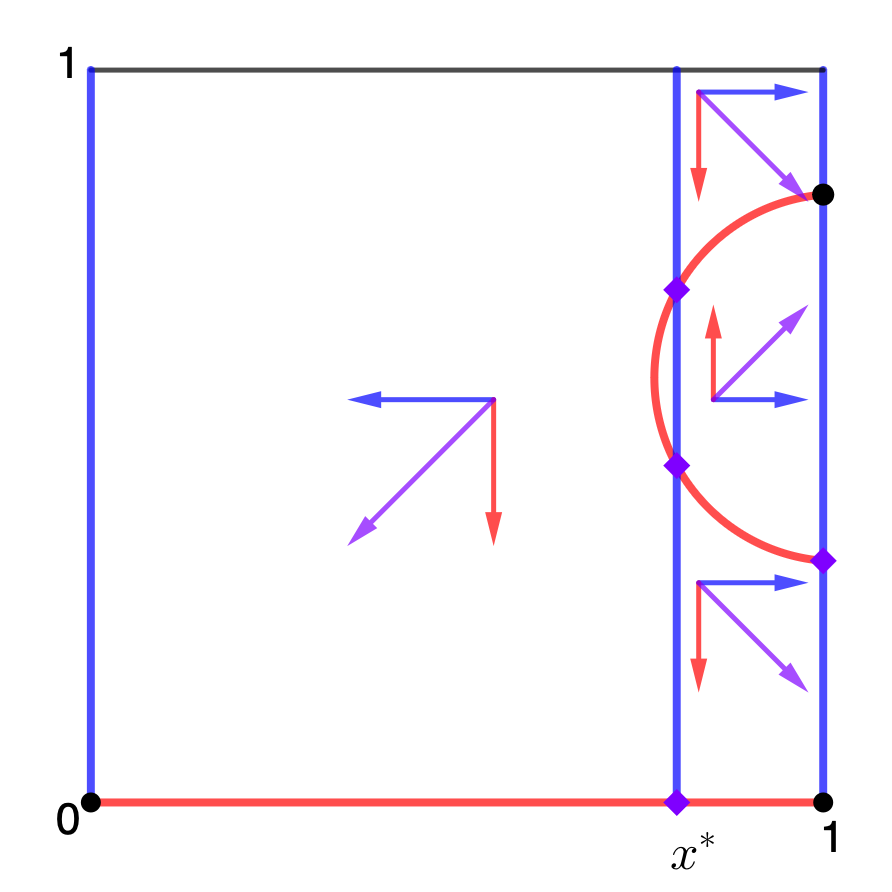}
         \caption{7 fixed points}
     \end{subfigure}
     \caption{Qualitative phase portrait}
     \label{phase}
\end{figure}

We give a qualitative phase portrait of the system when \eqref{root} has a well defined solution for $v$, which we will call $v^*(x,T_v)$. In Figure \ref{phase} (a), we have 6 fixed points and $v^*$ is on a red curve that moves smoothly when $T_v$ changes. When $T_v$ decreases, the red curve moves down. For example, when $T_v=15^\circ$C, the left endpoint of the curve will move below 0, leaving us 5 fixed points instead of 6, see Figure \ref{phase} (b). As the curve moves further down we will have 4 fixed points, when the curve only intersects $x=1$. When $T_v\le13^\circ$C, the curve will move out of the phase space, leaving us 3 fixed points, $(0,0),(1,0)$ and $(x^*,0)$. 

On the other hand when $T_v$ increases, instead of moving up, the red curve moves up and to the right showing a horizontal parabola that opens to the right. We will run into a 5-fixed-point case similar to phase portrait \ref{phase} (b). When $T_v=33^\circ$C, we see a horizontal parabola which gives us 7 fixed points, see Figure \ref{phase} (c). When $T_v\ge34^\circ$C the red curve moves out of our $[0,1]\times[0,1]$ phase space, leaving us with the 3 fixed points on the horizontal axis, $(0,0),(1,0)$ and $(x^*,0)$.

In phase portrait \ref{phase} (a), we can see that only two of the fixed points are stable, namely $(0,v^*(0,T_v))$ and $(1,v^*(1,T_v))$. All the other fixed points are unstable. When $T_v$ decreases and the red curve moves down, fixed point $(0,v^*(0,T_v))$ disappears and $(0,0)$ becomes stable. This stays true until $(1,v^*(1,T_v))$ disappears and $(1,0)$ becomes stable. 

Now when $T_v$ increases, the red curve becomes a parabola, starting somewhere between 0 and $x^*$ on the $x-$axis. It goes up and then right, still giving us 5 fixed points. Although the shape of the curve is different from phase portrait \ref{phase} (b), the stability analysis stays the same, so $(0,0)$ replaces $(0,v^*(0,T_v))$ as the stable fixed point. As the parabola moves to the right, we will have 6 fixed points again, where the parabola starts from somewhere between $x^*$ and 1 on the $x-$axis and moves left and up, then right and up. Although the number of fixed points changes and the phase portrait looks different, the stable fixed points are the same as before, $(0,0)$ and $(1,v^*(1,T_v))$. As $T_v$ increases further, we will reach phase portrait \ref{phase} (c) with 7 fixed points, and we will have one extra stable fixed point $(1,0)$. As the parabola moves further right, it will only intersect $x=1$, we have 5 fixed points and the same three stable ones, $(0,0),(1,0)$ and $(1,v^*(1,T_v))$. Finally when the red curve disappears on the right, we only have the three fixed points on the $x-$axis left and $(0,0)$ and $(1,0)$ are both stable.

Also as we pointed out earlier, when $T_c\ge2^\circ$C, $x^*$ is very close to 1. This means the right strip in the phase portraits in Figure \ref{phase} can be very thin.

\section{Simulation results}

We use the original uncoupled forest dieback model \eqref{forest} to compute the temperature of the forest for the first 10 years. Then we use a delay differential equation solver to find the evolution of the coupled social-climate model \eqref{social-couple} over years 10 to 200. We repeat the simulation with different initial conditions $x(0),v(0)$ and different $T_v$ and $T_c$ values. 

We already discussed the stability of the qualitative phase portraits of our system and the stable fixed points were $(0,0),(1,0),(0,v^*(0,T_v))$ and $(1,v^*(1,T_v))$. This tells us the proportion of mitigators should converge to either $1$ or $0$, which is what we see in our simulation results. In the simulation results, the vegetation coverage either stabilizes at a positive value or converges to zero, with some rare cases where it's oscillating over a long time. We should point out that some oscillations die out in 50 years or so and converge, so it's possible the oscillations that persist by year 200 could 

\begin{figure}[H]
    \vspace{-40pt}
     \centering
     \begin{subfigure}[b]{0.45\textwidth}
         \includegraphics[scale=0.45]{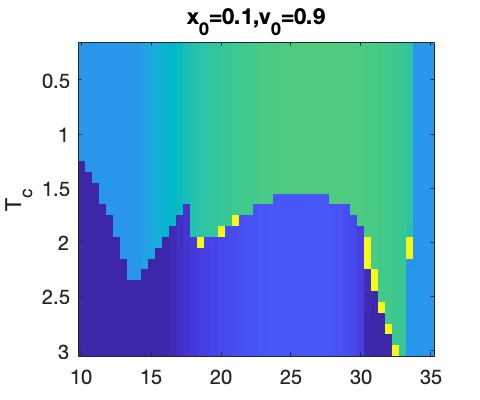}
     \end{subfigure}
     \qquad
     \begin{subfigure}[b]{0.45\textwidth}
         \includegraphics[scale=0.45]{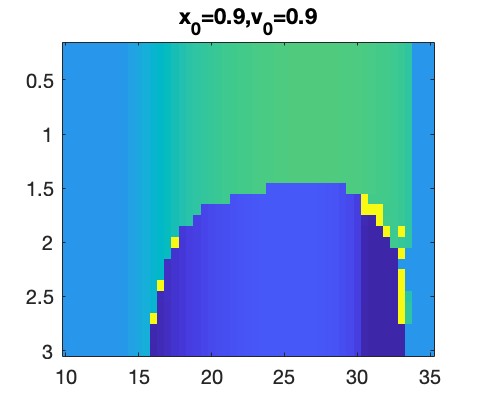}
     \end{subfigure}
     \\
     \vspace{-5pt}
     \begin{subfigure}[b]{0.45\textwidth}
         \includegraphics[scale=0.45]{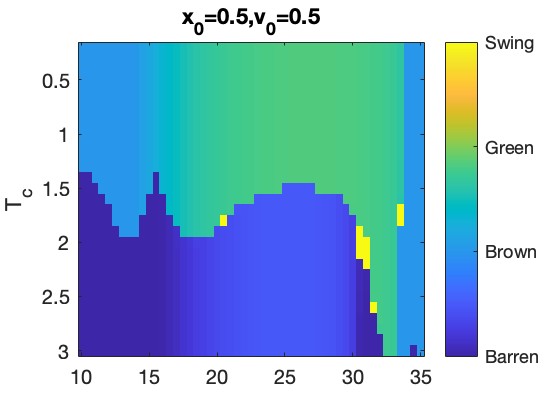}
     \end{subfigure}
     \\
     \vspace{-5pt}
     \begin{subfigure}[b]{0.45\textwidth}
         \includegraphics[scale=0.45]{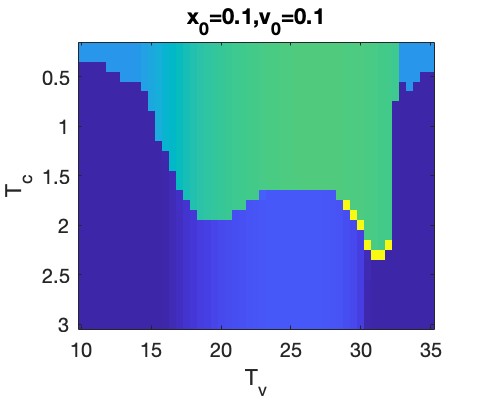}
     \end{subfigure}
     \qquad
     \begin{subfigure}[b]{0.45\textwidth}
         \includegraphics[scale=0.45]{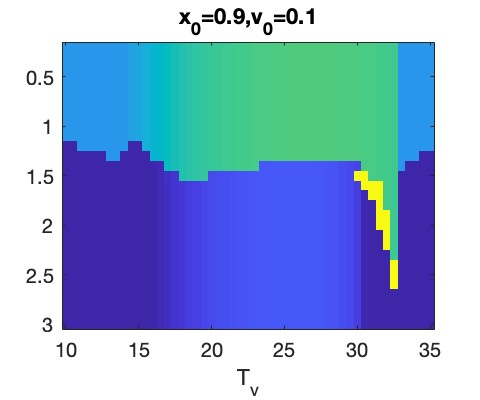}
     \end{subfigure}
     \caption{Forest coverage and mitigation after 200 years}
     \label{heatmap}
\end{figure}
\noindent die out if we computed for a longer time period as well. We only computed the model up to year 200 since it might not be realistic to compute it for longer considering the ambient temperature probably would have changed.

We call the cases where both $x$ and $v$ converge to 0 by year 200 "Barren" and the cases where $x$ converges to 1 and $v$ converges to 0 "Brown". "Green" represents cases where $x$ converges to 1 and $v$ converges to some positive value typically around 0.7 or 0.8. We will call the long term oscillations "Swing". Our simulation results are shown in Figure \ref{heatmap}.

Here the dark blue or purple region, the "Barren", means the forest dies out and nobody is a mitigator. The blue region, the "Brown", represents situations where the forest dies even though everyone is a mitigator. The green regions, the "Green", means the forest survives and everyone is a mitigator. The yellow region, the "Swing", represents cases where the vegetation and mitigation oscillate and don't converge by year 200. 

In general, we can view $T_c$ as the sensitivity level of the population towards warming. As we can see, typically, there is more "Green" for low values of $T_c$, which means people will adopt mitigative actions as soon as they feel a small amount of warming. As $T_c$ increases, people become less sensitive towards warming and don't adopt mitigation until the warming is rather severe. This might be too late as we can see a lot more "Barren" at high values of $T_c$.

The ambient temperature $T_v$ plays a role as well. When $T_v$ is either too high or too low, we also see more "Barren" and "Brown". This agrees with the general consensus that there is a critical warming that we cannot go back from. And of course if the ambient temperature is too low we probably can't survive either. It is worth noting that between around 20$^\circ$C and 30$^\circ$C, we see a lighter violet region, which represents no mitigators and the vegetation coverage converges to a low positive value, around 20\% for example. This happens when we have favorable environment where the forest stays even though there are no mitigators. 

The initial conditions also change the final result. For example, when we start from a high forest coverage of 90\% and low mitigation at 10\%, we have a good amount of "Green" and small amounts of "Barren". If the mitigation we start with is 90\%, the "Barren" region above 30$^\circ$C actually becomes bigger, this could be because we don't have a lot of room for improvement in behavior since the mitigation we started with is already high. And also if we start with low forest coverage at 10\%, we will have huge "Barren" regions.

We can also see that oscillations happen typically when we cross over from one region to the other. Our model is nonlinear and has time-delay, it

\begin{figure}[H]
    \vspace{-40pt}
    \centering
    \begin{subfigure}[b]{\textwidth}
        \includegraphics[scale=0.45]{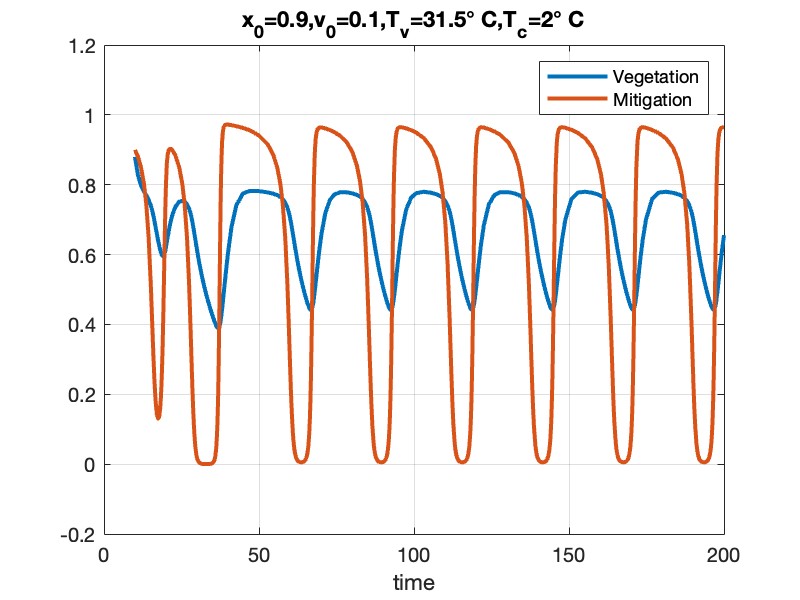}
    \end{subfigure}
    \\
    \begin{subfigure}[b]{\textwidth}
        \includegraphics[scale=0.45]{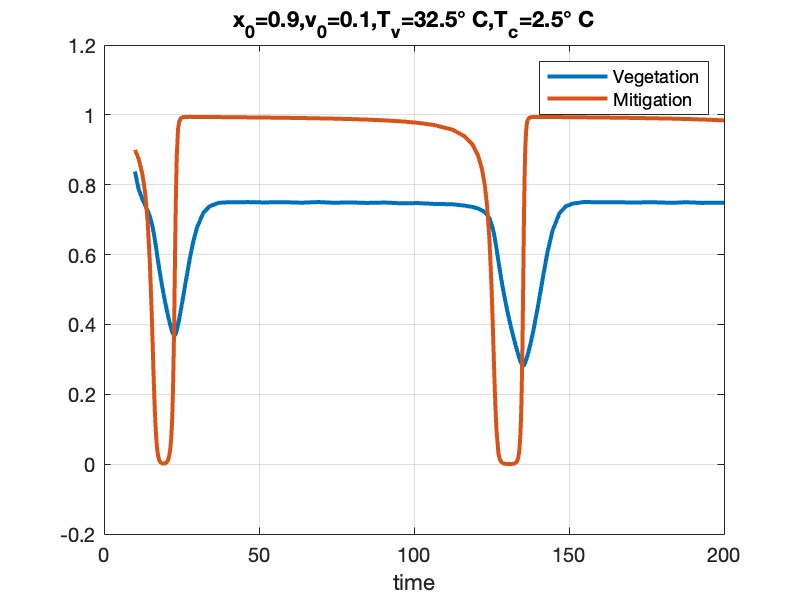}
    \end{subfigure}
    \caption{Long term oscillations}
    \label{osc}
\end{figure}

\noindent can give us very rich dynamics. Some oscillations die away within the first 50 years while some persist up to year 200. Some oscillations are small in amplitude and look like small persistent noise while others have value variation from 0 to 1, the full range of possible values. We show two specific plots for the evolution of vegetation and mitigation over 200 years for the initial condition $x(0)=0.9,v(0)=0.1$ in Figure \ref{osc}. As one can see, the mitigation varies from 0 to 1, the vegetation varies between 0.4 and 0.8, and the period can vary greatly as well.

\begin{figure}[H]
    \vspace{-40pt}
    \centering
    \begin{subfigure}[b]{\textwidth}
        \includegraphics[scale=0.45]{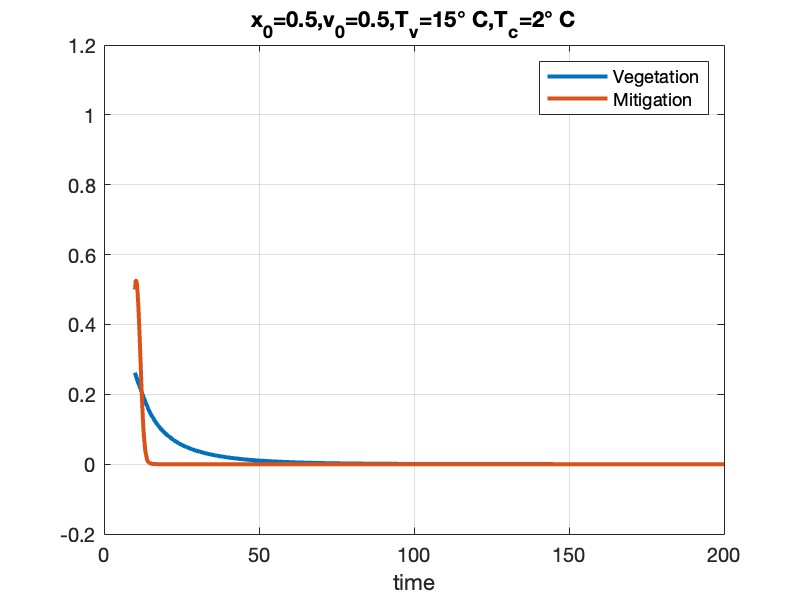}
    \end{subfigure}
    \\
    \begin{subfigure}[b]{\textwidth}
        \includegraphics[scale=0.45]{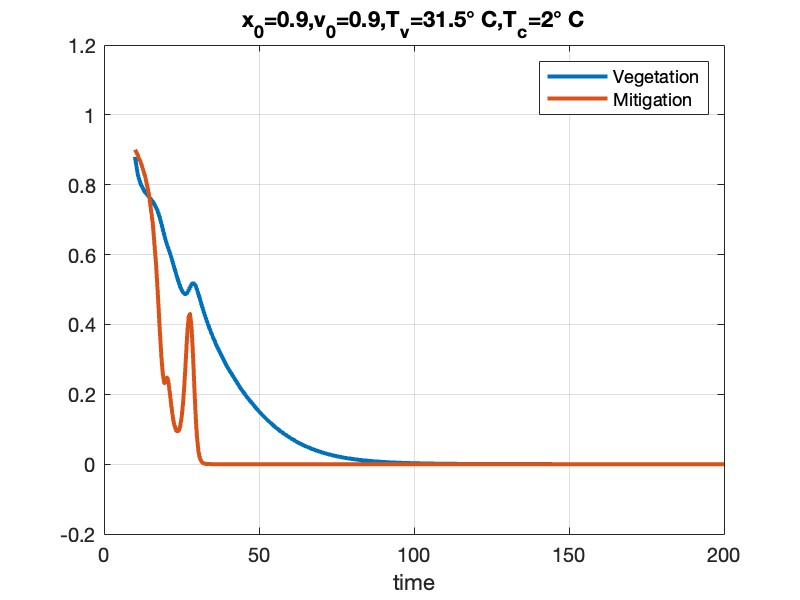}
    \end{subfigure}
    \caption{Two different instances of "Barren" evolution}
    \label{barren}
\end{figure}

\begin{figure}[H]
    \vspace{-40pt}
    \centering
    \begin{subfigure}[b]{\textwidth}
        \includegraphics[scale=0.45]{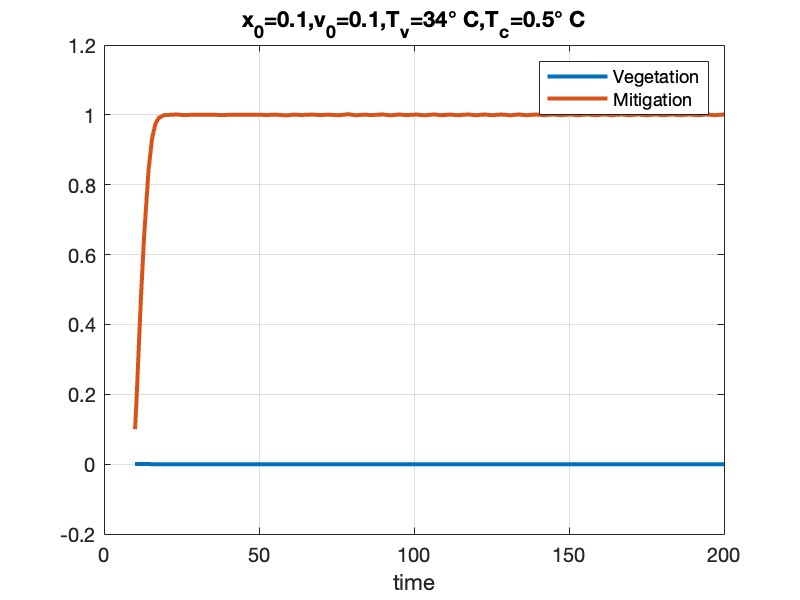}
    \end{subfigure}
    \\
    \begin{subfigure}[b]{\textwidth}
        \includegraphics[scale=0.45]{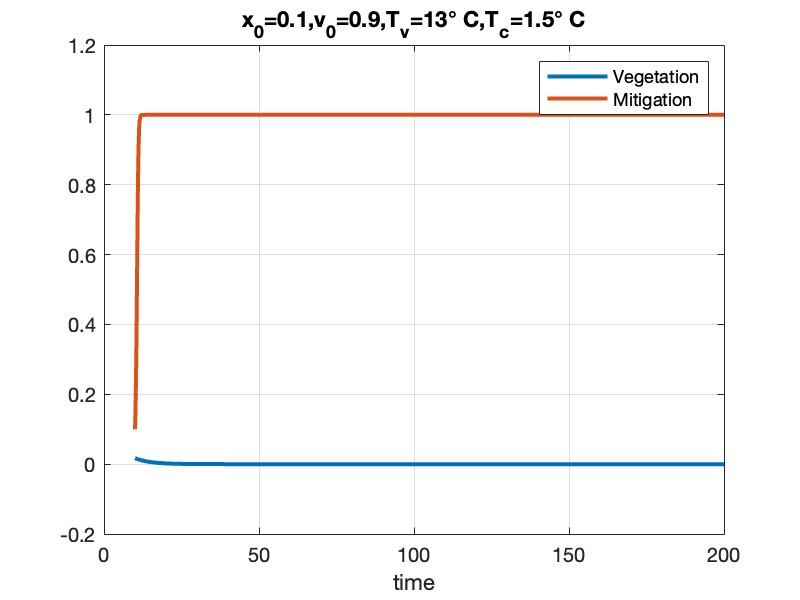}
    \end{subfigure}
    \caption{Two different instances of "Brown" evolution}
    \label{brown}
\end{figure}

\begin{figure}[H]
    \vspace{-40pt}
    \centering
    \begin{subfigure}[b]{\textwidth}
        \includegraphics[scale=0.45]{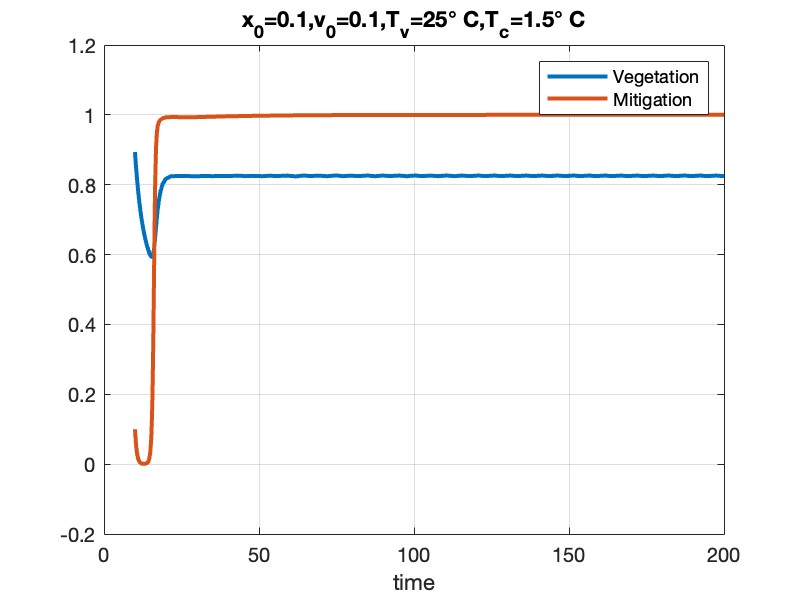}
    \end{subfigure}
    \\
    \begin{subfigure}[b]{\textwidth}
        \includegraphics[scale=0.45]{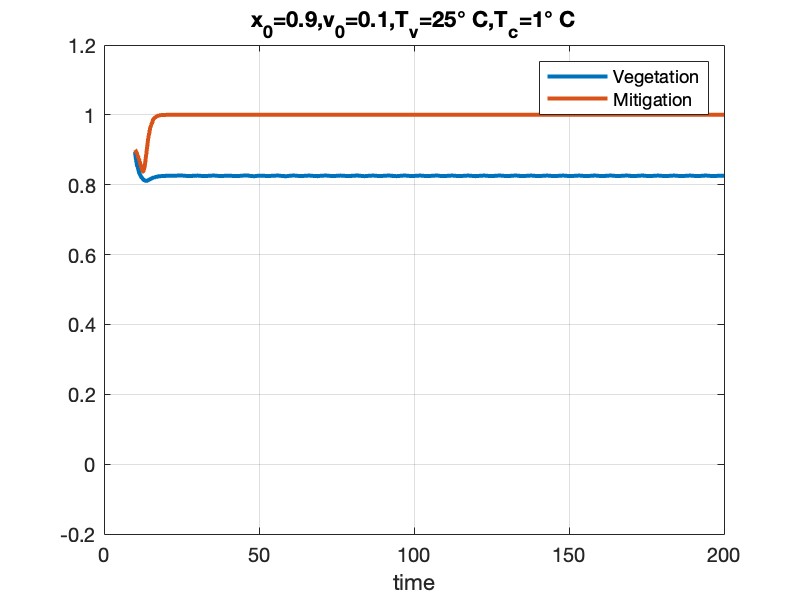}
    \end{subfigure}
    \caption{Two different instances of "Green" evolution}
    \label{green}
\end{figure}

\begin{figure}[H]
    \vspace{-40pt}
    \centering
    \begin{subfigure}[b]{\textwidth}
        \includegraphics[scale=0.45]{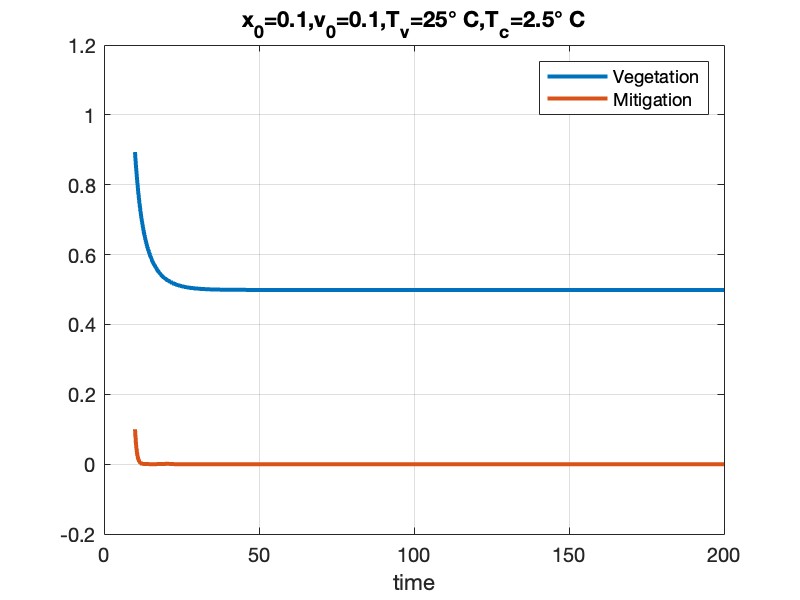}
    \end{subfigure}
    \\
    \begin{subfigure}[b]{\textwidth}
        \includegraphics[scale=0.45]{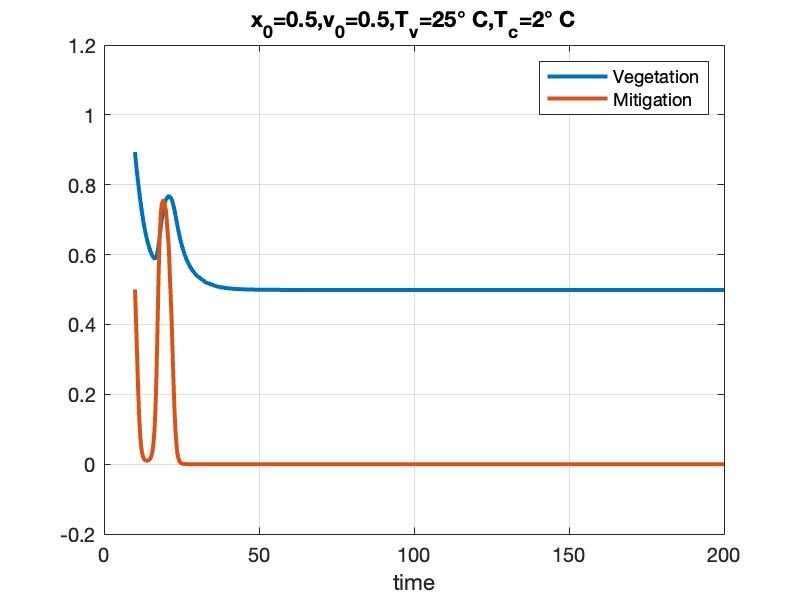}
    \end{subfigure}
    \caption{Two different instances of optimal ambient temperature evolution}
    \label{purple}
\end{figure}

We show two examples of evolution over time for "Barren" in Figure \ref{barren} and two examples of "Brown" in Figure \ref{brown}. Figure \ref{green} shows two cases of "Green" evolution and Figure \ref{purple} shows the cases where the ambient temperature is optimal and the forest survives even with no mitigators in the population at all.

\section{Discussion}

Evolutionary game theory is used in this study to build a time-delay system where the temperature of the forest depends on human behaviors and human behaviors also depend on the current temperature. The replicator equation determines the percentage of mitigators in the population based on the warming over the past 10 years. This means that the social part of the model has a time-delay effect, and the growth rate of the forest depends on the percentage of mitigators in the population. By studying this coupled social-climate model under changing parameter values, ambient temperature, or background temperature of the environment, and sensitivity to warming, we can gain insights into how human behavior affects climate change and how we can mitigate it.

Simulations of the coupled system over 200 years show us the varying results of forest dies out and no one is a mitigator, forest dies out and everyone is a mitigator, or the forest survives and everyone is a mitigator. There are rare cases where no one is a mitigator and yet the forest survives, but with low coverage. There are also rare oscillations where the proportion of mitigators varies between 0 and 1. These scenarios tell us that human behavior plays a crucial role in the survival of forests and that having everyone as a mitigator can help prevent forest dieback.

Our model was relatively simple and didn't consider the asymmetry of the population in aspects such as the contribution to global warming and resource inequalities \cite{menard21, wang2010effects,vasconcelos2014climate}. We could modify both the the climate part and the social part of our model to see how the asymmetries in the population affects the climate change. Our model has a deterministic process but there are stochastic game descriptions of the climate change as well \cite{barfuss2020caring}.

In sum, our work presents a transparent approach to modeling and analyzing the intricate interplay within social-climate systems. At its core, the climate model, grounded in forest dieback principles, may be straightforward, but it effectively incorporate the foundational effects of human behavior on both the climate and forest coverage. The decision-making process related to mitigation is firmly grounded on temperate forecasting and the sensitivity to  temperature change. Our model, while simple, exhibits rich dynamics. This resonates with Bob May's renowned work on logistic maps and chaos: ``Simple mathematical models with very complicated dynamics''~\cite{may1976simple}. Our current model and its extensions will help offer insights into the factors driving behavioral shifts in the context of climate dynamics.



 \bibliographystyle{elsarticle-num} 


\end{document}